\documentclass{article}

\usepackage{PRIMEarxiv}
\usepackage{tikz}
\usetikzlibrary{quantikz}
\usepackage{float}
\usepackage[utf8]{inputenc} 
\usepackage[T1]{fontenc}    
\usepackage{hyperref}       
\hypersetup{colorlinks,
 citecolor=blue,
 linkcolor=red,
 urlcolor = black}
\usepackage{algorithm}
\usepackage{algorithmic}
\usepackage{url}            
\usepackage{booktabs}       
\usepackage{amsfonts}       
\usepackage{amsmath, bm}
\DeclareMathOperator*{\argmin}{arg\,min}
\usepackage{nicefrac}       
\usepackage{microtype}      
\usepackage{lipsum}
\usepackage{fancyhdr}       
\usepackage{graphicx}       
\graphicspath{{media/}}     

\title{\textbf{One-Step Time Series Forecasting Using Variational Quantum Circuits}
}

\author{Payal Kaushik$^{1}$, Sayanatan Pramanik$^{2}$, M Girish Chandra$^{3}$, C V Sridhar$^{2}$\\
	\normalsize $^{1}$Indian Institute of Technology Dhanbad\\
	\normalsize $^{2}$TCS Incubation, India\\
	\normalsize $^{3}$TCS Research, India\\
	\normalsize \texttt{E-mail:kaushikpayal99@gmail.com, \{sayantan.pramanik, m.gchandra, sridhar.cv\}@tcs.com
	}
}

\begin{document}
\maketitle
\begin{abstract}
Time series forecasting has always been a thought-provoking topic in the field of machine learning. Machine learning scientists define a time series as a set of observations recorded over consistent time steps. And, time series forecasting is a way of analyzing the data and finding how variables change over time and hence, predicting the future value. Time is of great essence in this forecasting as it shows how the data coordinates over the dataset and the final result. It also requires a large dataset to ascertain the regularity and reliability. Quantum computers may prove to be a better option for perceiving the trends in the time series by exploiting quantum mechanical phenomena like superposition and entanglement. Here, we consider one-step time series forecasting using variational quantum circuits, and record observations for different datasets.
\end{abstract}


\section{Introduction}
Forecasting is a technique used to predict future values based on present and previous observations. There are various types of forecasting, such as qualitative techniques, time series analysis and projection, and causal models. Our primary concern in this paper is Time series forecasting. This forecasting type involves using one or more time series to make predictions \cite{chatfield2000time}. When one value is predicted ahead of time by utilizing the existing values of the time series, such forecasting is referred to as "One-step forecasting." Classically, various models are already in practice for this type of forecasting, such as the Autoregressive Integrated Moving Average (ARIMA) model. Time series forecasting is not a much-explored topic on quantum computers. Here, we have tried to implement one-step forecasting on quantum computers by transforming it into a regression problem and optimizing it using the Least-squares optimization method. A poster containing some results of this paper was accepted at the 2nd European Quantum Technologies Conference - EQTC 2021 \cite{Poster}. 
\subsection{Regression using Least squares optimisation}
    Regression is a very renowned problem in the field of machine learning. It is a statistical method that helps us analyze and understand the relationship between two or more variables of interest. In regression, we try to regress the value of the target variable with the help of independent variables \cite{montgomery2021introduction}.
    \begin{center}
		$y = f\left(x_1, x_2, ... , x_n\right)$
	\end{center}
	where $y$ is the target variable, and $x_1, x_2, ..., x_n$ are $n$ independent variables. When more than one independent variable is used and is varied linearly with the target variable, it is called Multiple Linear Regression.
	
	Mathematical formulation of Multiple Linear Regression-\\
	For a $m$ unit dataset $(y^i, x_1^i, x_2^i, ..., x_n^i)$ with $i$ varying from $1$ to $m$, assuming linear relationship between dependent and independent variables, the model takes the form-
	\begin{equation}\label{eq1}
		\hat{y} = \beta_0 + \beta_1 x_{1} + \beta_2x_{2}+ ... + \beta_n x_n 
	\end{equation}
	\begin{equation*}
	    y = \hat{y} + \epsilon
	\end{equation*}
	where $\hat{y}$ is the predicted target variable and $\epsilon$ is the error term. The above system can be written in matrix form as $Y = X\pmb{\beta} + \epsilon$, where 
	\begin{center}
    $X = \begin{bmatrix}
    	1 & x_{1}^1 & x_2^1 &... &x_{n}^1\\
    	1&x_{1}^2 & x_2^2 &... &x_{n}^2\\
    	... &... &... &...\\
    	1&x_{1}^m & x_2^m &... &x_{n}^m
    \end{bmatrix}$ \\
    \vspace{1em}
    $\beta = \begin{bmatrix}
    		\beta_0\\
    		\beta_1\\
    		...\\
    		\beta_n
    	\end{bmatrix}
    \, \, \, \, \hspace{1cm}  Y = \begin{bmatrix}
    	y^1\\
    	y^2\\
    	...\\
    	y^n
    \end{bmatrix}$
    \end{center}
	The regression hyperplane is found using the standard Least-squares method. This method minimizes the sum of the squares of the differences between the original and predicted value of the dependent variable.
	\begin{equation}
	    \min\left(\sum_{i=1}^m \left(\left(\hat{y}\right)^i - y^i\right)^2\right) 
	\end{equation}
	Differentiating the above equation with respect to parameters $\beta$, and simplifying, we get-
	\begin{equation}
		X^T X\beta = X^T Y
	\end{equation}
	Further, the above system of linear equations can be written and solved as follows-
	\begin{equation*}
		Ax = b
	\end{equation*}
	\begin{equation*}
		A := X^TX, \hspace{1cm} x := \beta, \hspace{1cm} b := X^T Y
	\end{equation*}
	\begin{equation*}
		x = A^{-1}b
	\end{equation*}
	Substituting the value of $x$ in place of $\beta$ in eq. (\ref{eq1}) gives the regression model.
\subsection{One-step forecasting as a Regression problem}
	It is the forecasting which predicts one value in the future by considering previous $m$ values. By treating one-step forecasting as a regression problem, we can predict the future value by regressing it using the earlier $m$ values in the following way-
	\begin{equation}
		\hat{y}_{t+1} = \sum^{m-1}_{k=0}c_k y_{t-k}
	\end{equation}
    where $\hat{y}_{t+1}$ is the value at the next time step, which is to be predicted, $y_t$, $y_{t-1}$, ...., $y_{t-m+1}$ are the previous $m$ values of time series, and $\epsilon$ is the error term. For the sake of simplicity, we replaced $t+1$ with $t$ in above equation.
    \begin{equation}\label{eq5}
		\hat{y}_{t} = c_1y_{t-1} + c_2y_{t-2}+ ... + c_{m-1}y_{t-m+1}
	\end{equation}
	\begin{equation}\label{eq_5}
		y_{t} = \hat{y}_t + \epsilon
	\end{equation}
	On comparing eq. (\ref{eq5}) with eq. (\ref{eq1}), it can be deduced that one-step forecasting can be treated as a regression problem.

\section{Algorithmic Steps}
\subsection{Classical Preprocessing}

    Considering the given time series $\{ y_1, y_2, ..., y_T\}$. It can be reframed into $ \{ y_{T-i}, y_{T-1-i}, y_{T-2-i}, ..., y_{T-m+1-i}\}_{i=0}^{T-m}$\\ where $m$ is the number of previous units to be considered for the prediction of the future value. Then, after appropriate scaling of data, it can be divided into two sets (say in the ratio of $k:n$ where $k+n=T$), we call them "Training set" and "Scaling set". Training set to be used for finding the relational variables $\{c_1, c_2, ..., c_m\}$ and the Scaling set for the scaling factor ($\lambda$).\\
    Further, by using the Training set, the equations can be prepared by incorporating the relation in eq. (\ref{eq5}) and eq. (\ref{eq_5}), and can be written in matrix form as-
    \begin{equation*}
		Y = X\beta + \epsilon
	\end{equation*}
    where
    \begin{center}
    $X = \begin{bmatrix}
    	y_{k-1} & y_{k-2} &... &y_{k-m+1}\\
    	y_{k-2} & y_{k-3} &... &y_{k-m}\\
    	... &... &... &...\\
    	y_{m-1} & y_{m-2} &... &y_{1}
    \end{bmatrix}$ \\
    \vspace{1em}
    $\beta = \begin{bmatrix}
    		c_1\\
    		c_2\\
    		...\\
    		c_{m-1}
    	\end{bmatrix}
    \, \, \, \, \hspace{1cm}  Y = \begin{bmatrix}
    	y_k\\
    	y_{k-1}\\
    	...\\
    	y_{m}
    \end{bmatrix}$
    \end{center}
    The system is transformed into the following matrix equation to solve using Least-squares optimization, as explained in the previous section. 
    \begin{equation*}
		Ax = b
	\end{equation*}
	where
	\begin{equation*}
		A = X^TX \hspace{1cm} x = \beta \hspace{1cm} b = X^TY
	\end{equation*}
    
\subsection{Quantum Preprocessing}
    
    The first step toward solving the system of linear equations on a quantum computer is to encode the problem in the quantum language. There are various ways to encode this equation for it to be used on a quantum computer, such as Basis Encoding, Amplitude Encoding, and Angle encoding \cite{weigold2020data}. The one used in this paper is Amplitude Encoding. The vectors $\Vec{b}$ and $\Vec{x}$ are first normalized and mapped to their respective quantum states $|b\rangle$ and $|x\rangle$. The mapping is such that the $i^{th}$ component of the vector $\Vec{b}$ corresponds to the amplitude of the $i^{th}$ basis state of the quantum state $|b\rangle$.
    \begin{equation*}
    	|b\rangle = \frac{\Vec{b}}{\|b\|_2} \hspace{2cm} |x\rangle = \frac{\Vec{x}}{\|x\|_2}
    \end{equation*}
    So, our problem can be restated using quantum states as follows-
    \begin{equation}\label{eq6}
    	A|x\rangle = |b\rangle
    \end{equation}
    The prepared matrix $A$ is now decomposed into a linear combination of unitaries ($A_1, A_2, ..., A_n$) with complex coefficients ($p_1, p_2, ..., p_n$), to be given as an input to the quantum algorithm-
    \begin{equation*}
    	A = \sum_n p_n A_n
    \end{equation*}
    Also, the vector $b$ is encoded into quantum state $|b\rangle$ using a unitary $U_b$ in the following way:
    \begin{equation*}
    	|b\rangle = U_b|0\rangle
    \end{equation*} 
    
\subsection{Solving system of Linear Equations}
    There are two ways to find the solution to the system of linear equations represented in eq. (\ref{eq6}). One is using pure quantum algorithms like HHL \cite{harrow2009quantum}, and the other is using Hybrid quantum-classical algorithms like Variational Quantum Linear Solver (VQLS) \cite{bravo2019variational}.
    
    HHL algorithm finds the solution to the system of linear equations purely on a quantum computer. It finds the solution in three steps- Quantum Phase Estimation (QPE), Ancilla Rotation, and Uncomputation. QPE is a quantum algorithm which, given a unitary $U_0$ with eigenvector $|u_j\rangle$ and
    eigenvalue $e^{i\lambda_j}$, will return $|\lambda_j\rangle|u_j\rangle$, where $|\lambda_j\rangle$ is the binary representation of $\lambda_j$. In the second step, it checks whether the QPE step was successfully able to estimate the eigenvalues. Finally, the values are retrieved in the Uncomputation step \cite{harrow2009quantum}.
    
    But quantum algorithms for solving linear systems of equations like HHL cannot be implemented in the near term due to the required high circuit depth. So, we have a Hybrid quantum-classical algorithm, called Variational Quantum Linear Solver (VQLS), for solving linear systems on near-term quantum computers.
\subsubsection{Variational Quantum Linear Solver (VQLS)}

    VQLS is a hybrid solution using both classical and quantum computing methods to solve the quantum system of equations. This algorithm is not an iteration on the HHL \cite{harrow2009quantum} but a proposed intermediary solution to HHLs high demand of qubits and high quality of computation. It is designed to work on so-called Noisy Intermediate-Scale Quantum computers (NISQ) \cite{bharti2021noisy} by reducing the depth of the quantum circuit needed to solve the problem. It does this essentially by moving parts of the algorithm back to a classical computer. It has been used as a subroutine for the forecasting problem. The main goal of this algorithm is to minimize the cost function using the classical minimizer. The cost function is defined in terms of the overlap between the quantum states $|b\rangle$ and $\frac{A|x\rangle}{\sqrt{\langle x|A^\dagger A|x\rangle}}$. To estimate this cost, we use an efficient quantum circuit. The $\alpha$ parameters of the quantum circuit are determined classically and fed to the quantum computer. The quantum computer then prepares the state $|x(\alpha)\rangle$ and, with it, efficiently estimates the cost function, which is then returned to the classical computer. After this, we used a classical minimizer to minimize the cost function. The new $\alpha$ is then fed again into the quantum computer, which again prepares the state $|x(\alpha)\rangle$. This loop repeats itself until the desired minimal cost is reached, i.e., $C(\alpha)<=\gamma$. The system then outputs the optimal $\alpha^*$ which can be used to prepare a state $|x\rangle$ using the Ansatz (trainable gate sequence $V(\alpha)$) \cite{bravo2019variational}, such that- 
    \begin{equation}\label{eq_8}
        |x\rangle = |x(\alpha^*)\rangle = V(\alpha^*)|0\rangle
    \end{equation}
    
    
    This algorithm takes the unitaries $A_1, A_2, ..., A_n$, and quantum state $|b\rangle$ as the input and gives $|x\rangle$ as the output, where
    \begin{equation}\label{eq_6}
    	|x\rangle = \frac{\Vec{x}}{\|x\|_2}
    \end{equation}
    Therefore, 
    \begin{equation}\label{eq7}
    	\Vec{x} = |x\rangle \|x\|_2
    \end{equation}
    Based on the previous discussion, $\Vec{x}$ should be equal to $\begin{bmatrix}
	c_1 & c_2 &... & c_{m-1} 
    \end{bmatrix}$ and $\|x\|_2 = \sqrt{c_1^2 + c_2^2 + .. + c_{m-1}^2}$.
    
    \textbf{Ansatz}
    
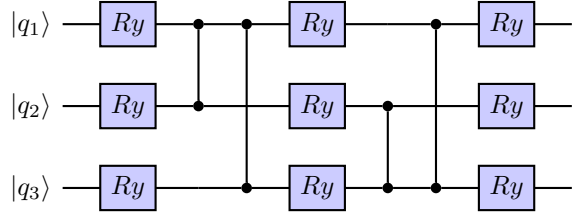
\begin{figure}[h!]
        \centering
        \begin{quantikz}
            \lstick{$\ket{q_1}$} & \gate[style={fill=blue!20}]{Ry} & \ctrl{1} & \ctrl{2} & \gate[style={fill=blue!20}]{Ry} & \qw& \ctrl{2}& \gate[style={fill=blue!20}]{Ry} & \qw\\
            \lstick{$\ket{q_2}$}&\gate[style={fill=blue!20}]{Ry} & \ctrl{} &\qw & \gate[style={fill=blue!20}]{Ry} &\ctrl{1} & \qw & \gate[style={fill=blue!20}]{Ry} &\qw\\
            \lstick{$\ket{q_3}$}&\gate[style={fill=blue!20}]{Ry} & \qw &\ctrl{} & \gate[style={fill=blue!20}]{Ry} & \ctrl{} & \ctrl{} & \gate[style={fill=blue!20}]{Ry} & \qw 
        \end{quantikz}
        \caption{Variational ansatz for 3 qubits}
        \label{fig:ansatz}
    \end{figure}

    The ansatz prepares a potential solution $|x(\alpha)\rangle = V (\alpha)|0\rangle$, where $V (\alpha)$ is a trainable gate sequence. The ansatz used here consists of layers of single-qubit $Ry$ rotations with embedded layers of $CZ$ rotations as shown in Fig. \ref{fig:ansatz}.
    \\
    \textbf{Cost Function}
    
    The goal of this algorithm is to minimize the cost function to find optimal $\alpha^*$. So when $|\psi\rangle = A|x(\alpha)\rangle$ ($|x(\alpha)\rangle$ is the state produced by ansatz for parameter $\alpha$) is very close to $|b\rangle$, the value of the cost function should be small, and vice versa when the vectors are orthogonal. The cost function used is shown below \cite{bravo2019variational}.
    \begin{equation*}
        C_{p} = \langle\psi|\psi\rangle - \langle \psi|b\rangle\langle b|\psi\rangle
    \end{equation*}
    The second term indicates the projection of $|\psi\rangle$ on $|b\rangle$. This term is subtracted from another number to get a small number when the inner product of $|\psi\rangle$ and $|b\rangle$ is greater, and the opposite for when they are close to being orthogonal. Further cost function is normalized to reduce the chances of deflection when norm of $|\psi\rangle$ is very less \cite{bravo2019variational}.
    \begin{equation*}
        \hat{C_{p}} = \frac{\langle\psi|\psi\rangle}{\langle\psi|\psi\rangle} - \frac{\langle \psi|b\rangle\langle b|\psi\rangle}{\langle\psi|\psi\rangle}
    \end{equation*}
    \begin{equation*}
        \hat{C_{p}} = 1 - \frac{|\langle b|\psi\rangle|^{2}}{\langle\psi|\psi\rangle}
    \end{equation*}
    To implement cost function, the above two terms need to be calculated, which are estimated using Hadamard Test \cite{aharonov2009polynomial} and Special Hadamard Test. We have worked out a few changes to handle dense marices (See Appendix \ref{app}). Special Hadamard Test is a controlled Hadamard test which is used to calculate the term $|\langle b|\psi\rangle|^{2}$ by controlling all the unitaries \cite{VQLS}.
\subsection{Finding scaling factor}

    The above-prepared solution state $|x\rangle$ in eq. (\ref{eq_8}) will be used for forecasting, but to retrieve the predicted result for further use, we require a scaling factor that should work as the norm of the predicted outcome. From eq. (\ref{eq7}), it is evident that to get the value of $\Vec{x}$, we need to multiply the obtained state with its norm. But, it is not possible to estimate the normalization factor in quantum computing. So, a scaling factor ($\lambda$) is calculated to estimate the value of $\Vec{x}$.
    
    And to do so, the scaling set of the dataset is used. The inner product between $\Vec{x}$ and $\Vec{y}$ where $\Vec{x} = \begin{bmatrix} c_1 & c_2 &... & c_{m-1} \end{bmatrix}$ and $\Vec{y} = \begin{bmatrix} y_{j-1} & y_{j-2} &... & y_{j-m+1} \end{bmatrix}$ gives $y_{j}$. A quantum subroutine called SWAP TEST as shown in Fig. \ref{fig:swap}, is used between $|x\rangle$ and $|y\rangle$ to execute the inner product \cite{barenco1997stabilization}.\\
    \begin{figure}[h!]
        \centering
        \begin{quantikz}
            \lstick{$\ket{0}$} & \gate[style={fill=purple!25}]{H} & \ctrl{2} & \gate[style={fill=purple!25}]{H} & \qw & \meter[style={fill=yellow!25}]{} & \qw\\
            \lstick{$\ket{x}$}&\qw & \targX{} &\qw & \qw &\qw & \qw \\
            \lstick{$\ket{y}$}&\qw & \targX{} &\qw & \qw & \qw & \qw
        \end{quantikz}
        \caption{SWAP Test}
        \label{fig:swap}
    \end{figure}
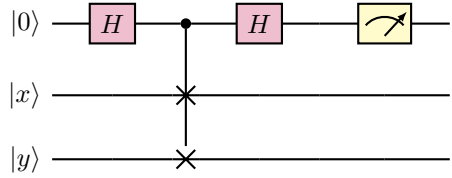\\
    The inner product has been calculated from the eq. (\ref{eq9}) [$P(0)$ is the probability of measuring $0$ on the controlling qubit]:
    \begin{equation}\label{eq9}
    	P(0) = \frac{1}{2} + \frac{1}{2}\langle x|y\rangle^2
    \end{equation}
    Now, the scaling factor $\lambda$ can be estimated by applying Least-squares method on the scaling dataset.
    \begin{equation*}
    	\lambda^* = \argmin_{\lambda} \left(\sum_{j=1}^{n} (y_j - \lambda\|y\|_2 \langle y|x\rangle)^2\right)
    \end{equation*}
    
\subsection{Forecasting}

    Forecasting is the process of predicting the future value using the existing values. The final quantum prediction has been made by performing the SWAP TEST on the quantum state $|x\rangle$ and $|y_1\rangle$, where $|x\rangle$ is the solution state prepared using the VQLS algorithm and $|y_1\rangle$ is the quantum state representing $\Vec{y_1}$, where $\Vec{y_1} = \begin{bmatrix}
    	y_T & y_{T-1} & y_{T-2} &... & y_{T-m+1}
    \end{bmatrix}$. So,
    \begin{equation*}
    	|y_1\rangle = \frac{\Vec{y_1}}{\|y_1\|_2} 
    \end{equation*}
    The quantum state $|y_1\rangle$ is prepared by applying unitary $U_1$ to $|0\rangle^{\otimes p}$, where $p = \log_2 m$, such that,
    \begin{equation*}
    	|y_1\rangle = U_1|0\rangle^{\otimes p}
    \end{equation*}
    The solution state $|x\rangle$ is prepared by passing the optimized parameters $\alpha^{*}$, found from the Variational Quantum Linear Solver algorithm which involves classical optimizer like COBYLA, to the variational ansatz.
    \begin{equation*}
    	|x\rangle = V(\alpha^*)|0\rangle 
    \end{equation*}
    Final prediction $\hat{y}_{T+1}$, is made by performing SWAP TEST on the states $|x\rangle$ and $|y_1\rangle$, estimating the inner product, and performing necessary scaling.
    \begin{equation*}
    	\hat{y}_{T+1} = \lambda(\|y_1\|_2) \langle x|y_1\rangle
    \end{equation*}
    
    \begin{figure}[h!]
    	\centering
    	\includegraphics[scale=0.5]{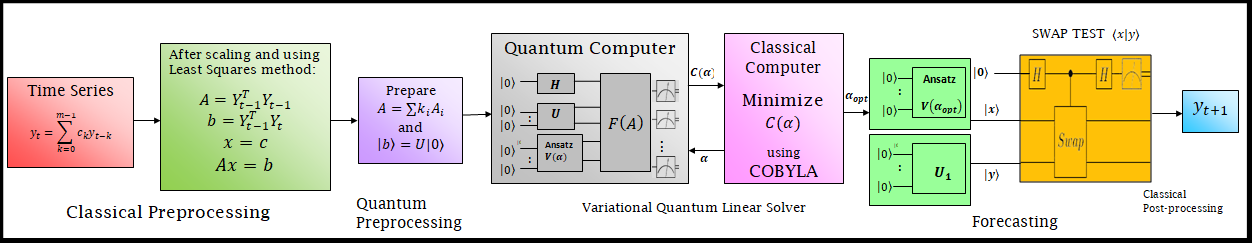}
    	\caption{Schematic diagram for the algorithm.}
    	\label{fig:algo}
    \end{figure}
\section{Datasets}\label{data}

    The algorithm has been numerically experimented with various real-life datasets like Historical data from the Indian market for the index National Stock Exchange Fifty (NIFTY) for May of the year 2021 \cite{NIFTY}. It has also been implemented for exemplary stocks data available for prediction and analysis \cite{Stocks_3M}. We have also used the elecequip (Electrical Equipment manufactured in the Euro Area) dataset \cite{Elecequip}. This data shows the number of new orders for electrical equipment in the Eurozone area. 
    Furthermore, the algorithm has been executed on yearly Revenue and profit data for Alphabet \cite{Google} and GigaMedia \cite{GIGM}, and quarterly revenue data for Alphabet \cite{Google}. 
    We have also performed the algorithm on the Stocks data of the giant computer company IBM (International Business Machines) \cite{IBM}. Lastly, we have also analyzed and tried to forecast the rainfall for the months of January and June from Indian Meteorological data \cite{Rainfall}.

\section{Implementation}
    We have used Qiskit \cite{wille2019ibm} for quantum part of the variational algorithm and NumPy \cite{oliphant2006guide}, Pandas \cite{mckinney2011pandas}, and SciPy \cite{virtanen2020scipy} for classical part. The cost function has been minimized using classical minimizer COBYLA.
    
\section{Results and Conclusion}

    The quantum algorithm has been carried out for $m=2$ and $m=4$. For all the datasets, $m=2$ gave better results than $m=4$. For Data1, i.e., Data from the Indian Market for the index National Stock Exchange Fifty (NIFTY), the values for two different values of $m$ are shown in the following table.
    \begin{center}
    \scalebox{1.2}{%
        \begin{tabular}{|c|c|c|c|c|c|}
            \hline
            $m$  & \textbf{Actual} & \textbf{Classical}& \textbf{Quantum} & \textbf{Classical} & \textbf{Quantum}\\
            & \textbf{value} & \textbf{Prediction} & \textbf{Prediction} & \textbf{Error} & \textbf{Error}\\
            \hline
            \textbf{2} & 15435.65 & 15368.761 & 15441.40 & 0.434\% & 0.037\% \\
            \textbf{4} & 15435.65 & 15366.89 & 15384.92 & 0.445\% & 0.328\% \\
            \hline
        \end{tabular}  }  
    \end{center}
    The simulation has been performed for datasets mentioned in Sec. \ref{data} for $m=2$.
    \begin{equation*}
    	y_{t+1} = c_0y_t + c_1y_{t-1}
    \end{equation*}
    \begin{figure}[h!]
        \centering
        \fbox{\includegraphics[scale=0.4]{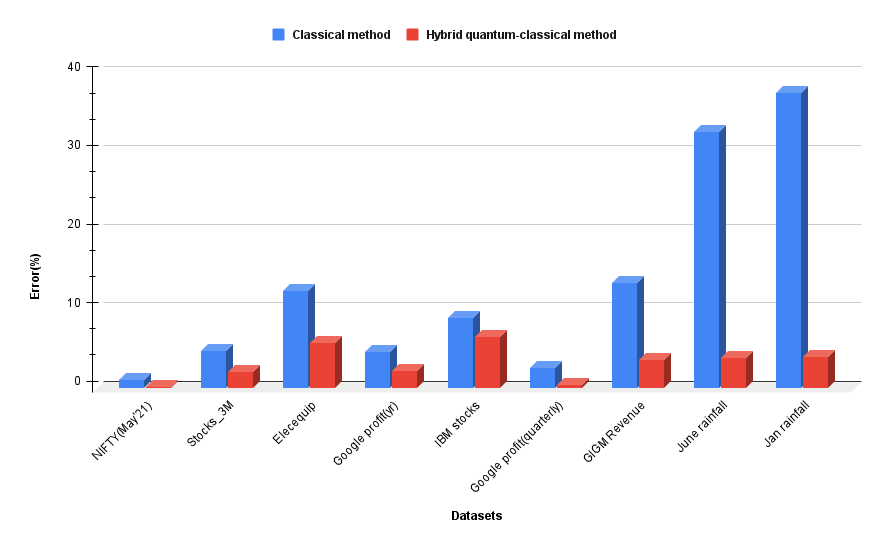}}
        \caption{Chart showing errors for the algorithm performed using classical method and hybrid quantum-classical method.}
        \label{fig:error}
    \end{figure}\\
    From the chart in the Fig. \ref{fig:error}, it can be concluded that the hybrid quantum-classical algorithm performs better than the purely classical algorithm for all tested datasets. Of course, in this preliminary study we have considered simple classical linear regressor and small windows for the "previous data". The whole intention is to demonstrate the applicability of the quantum-enhanced algorithms for simple one-step forecasting. The improved performance of the hybrid quantum-classical strategy can be attributed to the better exploitation of the correlation among the previous samples in the quantum representation.
    
\bibliographystyle{unsrt}  
\bibliography{references} 
\appendix
\section{\centering Algorithm to evaluate $\pmb{\langle\psi|\psi\rangle}$}\label{app}
    
    To evaluate the cost function, we need to calculate the value of $\langle\psi|\psi\rangle$. But it's not straightforward because $|\psi\rangle = A|x(\alpha)\rangle$ where $A$ is not a unitary matrix. It means $|\psi\rangle$ is not a unit vector due to which $\langle\psi|\psi\rangle \neq 1$. So, we have used Hadamard test to calculate the value of $\langle\psi|\psi\rangle$.\\ If we have some unitary $U$ and some state $|\phi\rangle$, then the subroutine called Hadamard Test, finds the expectation value $\langle \phi |U| \phi \rangle$ of $U$ with respect to the state $|\phi\rangle$. The circuit in Fig. \ref{fig:had} evaluates the desired value.
    \begin{figure}[h!]
        \centering
        \begin{quantikz}
            \lstick{$\ket{0}$} & \gate[style={fill=purple!25}]{H} & \ctrl{1} & \gate[style={fill=purple!25}]{H} & \qw & \meter[style={fill=yellow!25}]{} & \qw\\
            \lstick{$\ket{\phi}$}&\qw & \gate[style={fill=blue!25}]{U} &\qw & \qw &\qw & \qw
        \end{quantikz}
        \caption{Hadamard Test}
        \label{fig:had}
    \end{figure}
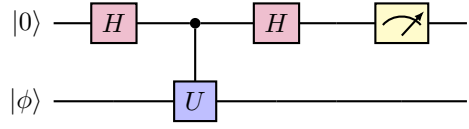\\
    Probability of measuring first qubit as $0$ and 1 are-
    \begin{center}$P(0) = \frac{1}{2}(1+ Re\langle\phi|U|\phi\rangle)$ \hspace{1cm}
    $P(1) = \frac{1}{2}(1 - Re\langle\phi|U|\phi\rangle)$\end{center}
    By taking the difference, we have $P(0) - P(1) = Re\langle\phi|U|\phi\rangle$. Now, to evaluate $\langle\psi|\psi\rangle$, we can simplify and write-
    \begin{equation*}
        \langle\psi|\psi\rangle = \langle x(\alpha)|A^{\dagger}A|x(\alpha)\rangle
    \end{equation*}
    where $A = \sum_n p_n A_n$, so we can write $\langle\psi|\psi\rangle = \sum_m\sum_n p_m^*p_n \langle x(\alpha)|A^{\dagger}_mA_n|x(\alpha)\rangle$, where $|x(\alpha)\rangle = V(\alpha)|0\rangle$. 
    \begin{algorithm}
    \caption{Pseudocode of our implementation of our algorithm to evaluate $\langle\psi|\psi\rangle$}
    \begin{algorithmic}
        \REQUIRE $V(\alpha)$ a parameterized quantum circuit, complex coefficients $p_1, p_2, ..., p_n$ and unitaries $A_1, A_2, ..., A_n$, and $m=n$
        \FOR{$i = 1,2,3,...,m$}
        \FOR{$j = 1,2,3,...,n$}
        \STATE $k \leftarrow p_i^*p_j$\\
        \STATE Add Hadamard gate to auxillary qubit to prepare the quantum circuit for Hadamard test
        \STATE Prepare quantum state $|x(\alpha)\rangle$ using Variational Ansatz $V(\alpha)$\\
        \STATE $|x(\alpha)\rangle = V(\alpha)|0\rangle$\\
        Add Hadamard gate to auxillary qubit in the circuit\\
        \IF{$A_i = X$}
        \STATE Apply $CX$ rotation on state $|x(\alpha)\rangle$ controlled by auxillary qubit\\
        \ELSIF{$A_i = Y$}
        \STATE Apply $CY$ rotation on state $|x(\alpha)\rangle$ controlled by auxillary qubit\\
        \ELSIF{$A_i = Z$}
        \STATE Apply $CZ$ rotation on state $|x(\alpha)\rangle$ controlled by auxillary qubit\\
        \ENDIF
        \STATE Repeat the process for $A_j$ 
        \STATE Add Hadamard gate to auxillary qubit in the circuit
        \STATE Measure the state of auxillary qubit and calculate the probabilities $P(0)$ and $P(1)$.
        \STATE $sum \leftarrow sum + k(P(0)-P(1))$
        \ENDFOR
        \ENDFOR
        \RETURN Value of $\langle\psi|\psi\rangle$ as $sum$
    \end{algorithmic}
    \end{algorithm}
\end{document}